\documentclass[12pt]{iopart}
\usepackage{braket}
\usepackage{amssymb}
\usepackage{xcolor}
\usepackage{graphicx}
\usepackage{hyperref}
\usepackage{notoccite}

\newcommand{\ham}{H(\vec{V})}

\newcommand{\be}{\begin{equation}}
\newcommand{\ee}{\end{equation}}
\newcommand{\VXsym}{V_{X,\textrm{\tiny sym}}}
\newcommand{\Ngates}{N_{\textrm{\tiny gates}}}

\begin{document}
\title[RX in TQDs for Spin-Photon Transduction]{Resonant Exchange Operation in Triple-Quantum-Dot Qubits for Spin-Photon Transduction}
\author{Andrew Pan, Tyler E. Keating, Mark F. Gyure, Emily J. Pritchett\footnote{Present address: IBM Thomas J Watson Research Center, 1101 Kitchawan Road, Yorktown Heights, NY 10598}, Samuel Quinn, Richard S. Ross, Thaddeus D. Ladd, Joseph Kerckhoff}
\address{HRL Laboratories, LLC, 3011 Malibu Canyon Road, Malibu, California 90265, USA}
\ead{aspan@hrl.com}

\begin{abstract}
Triple quantum dots (TQDs) are promising semiconductor spin qubits because of their all-electrical control via fast, tunable exchange interactions and immunity to global magnetic fluctuations. 
These qubits can experience strong transverse interaction with photons in the resonant exchange (RX) regime, when exchange is simultaneously active on both qubit axes. 
However, most theoretical work has been based on phenomenological Fermi-Hubbard models, which may not fully capture the complexity of the qubit spin-charge states in this regime. 
Here we investigate exchange in Si/SiGe and GaAs TQDs using full configuration interaction (FCI) calculations which better describe practical device operation. 
We show that high exchange operation in general, and the RX regime in particular, can differ significantly from simple models, presenting new challenges and opportunities for spin-photon coupling. 
We highlight the impact of device electrostatics and effective mass on exchange and identify a new operating point (XRX) where strong spin-photon coupling is most likely to occur in Si/SiGe TQDs. 
Based on our numerical results, we analyze the feasibility of a remote entanglement cavity iSWAP protocol and discuss design pathways for improving fidelity.
Our analysis provides insight into the requirements for TQD spin-photon transduction and  demonstrates more generally the necessity of accurate modeling of exchange in spin qubits.
\end{abstract}

\section{\label{intro}Introduction}
Semiconductor spins are of great interest for quantum computing because of their compactness, long-lived coherence, and potential scalability \cite{ladd_quantum_2010}. 
Out of the many possible methods for spin manipulation, the exchange interaction is particularly promising because it is fast, tunable, and electrically controllable. 
Exchange can be used as the sole control mechanism for triple-quantum-dot (TQD) encoded qubits, which reside within a decoherence-free subsystem (DFS) protected against global magnetic field fluctuations \cite{divincenzo_universal_2000,laird_coherent_2010,fong_universal_2011}. 
High-fidelity Si/SiGe TQD qubit operation has been shown experimentally \cite{andrews_quantifying_2019}, which is particularly relevant given the known physical (\textit{e.g.}, nuclear-spin-free isotope) and technological (mature, scalable processing) advantages of silicon. 
However, transporting quantum information between distantly situated TQDs will require a mechanism different from exchange, due to the latter's intrinsically short-range nature.

Microwave (MW) photons in superconducting resonators are an attractive choice to mediate long-range spin entanglement by leveraging the powerful, highly successful techniques developed in circuit QED \cite{childress_mesoscopic_2004,burkard_superconductor-semiconductor_2019}. 
The fundamental challenge is to engineer sufficient spin-photon interaction: the small, fixed electron spin magnetic moment couples very weakly to the magnetic field of a MW photon \cite{schoelkopf_wiring_2008}, while spin qubits typically have very weak (but tunable) charge dipoles which suppress decoherence but also electrical interaction with photons. 
Efficient transduction requires fine-tuning a finite spin-charge moment for photon coupling while minimizing charge noise susceptibility. 
In a weak spin-orbit material like silicon, a simple way to engineer such an interaction is to use magnetic field gradients \cite{hu_strong_2012,beaudoin_coupling_2016,benito_input-output_2017}, for instance from an on-chip micromagnet.
This method has been used to demonstrate strong spin-photon coupling \cite{mi_coherent_2018,samkharadze_strong_2018} and cavity-mediated spin-spin interactions \cite{borjans_resonant_2020} in silicon double-quantum-dots (DQDs). 
However, field gradients are undesirable for DFS TQDs since they induce qubit leakage \cite{andrews_quantifying_2019}. 
Fortunately, the multiple control axes for TQDs allow for other ways to achieve strong coupling \cite{russ_three-electron_2017} such as the resonant exchange (RX) regime \cite{taylor_electrically_2013,medford_quantum-dot-based_2013}, which hosts spin-photon ``sweet spots'' where a vanishing longitudinal (DC) electric dipole suppresses charge noise and exists alongside a transverse (AC) dipole for MW coupling \cite{russ_asymmetric_2015,russ_long_2015,russ_coupling_2016,srinivasa_entangling_2016}. 
Following these guidelines, strong coupling has been shown between GaAs TQDs using RX to a superconducting resonator \cite{landig_coherent_2018} and transmon \cite{landig_virtual-photon-mediated_2019}, respectively.

While these results validate the general concept of RX operation, better device modeling is needed for detailed performance assessments. 
Most theoretical device-level analyses of TQD qubits in the literature use phenomenological Fermi-Hubbard (FH) models, \textit{e.g.}, \cite{russ_three-electron_2017,taylor_electrically_2013,srinivasa_entangling_2016}. 
While useful for qualitative intuition, these models, at least as commonly used, cannot fully capture the dependence of exchange on device electrostatics. 
Such details matter since RX operation requires particular biasing configurations to couple the appropriate charge states. 
Accurate descriptions of the relevant physics require directly solving the multi-electron Schrodinger equation, for instance using the full configuration interaction (FCI) method \cite{szabo_modern_1996}. 
In this paper we present a numerical FCI-based analysis of RX in realistic TQDs, focusing on the implications for spin-photon transduction. 
In Section \ref{fhtqd} we describe the basic elements of RX control in TQDs and the implementation and predictions of the FH model in this regime. 
Section \ref{operation} focuses on the evaluation of TQD exchange using FCI calculations. 
We explain in detail how device geometry and electrostatics lead to behavior not predicted by simpler FH models and show that resonant exchange in typical Si/SiGe devices primarily occurs in an unexpected charge state, which we will denote as XRX. 
In Section \ref{spinphoton} we discuss the feasibility of spin-photon coupling based on these findings and apply them to predict the performance of cavity-mediated spin-spin entanglement. 
We also compare our numerical results with experimental data in GaAs and discuss device design prospects for improving spin-photon coupling.
We conclude in Section \ref{conclude}. 
The appendices provide more detail on our models for $g$ and charge noise, numerical computations of noise multipoles and detuning dependence of exchange, and estimation of protocol fidelity for different noise types.

\section{\label{fhtqd}TQD Operation and Fermi-Hubbard Predictions}
\subsection{TQD Qubit Operation and Key Quantities for Spin-Photon Coupling}
\begin{figure}
	\includegraphics[width=1\textwidth]{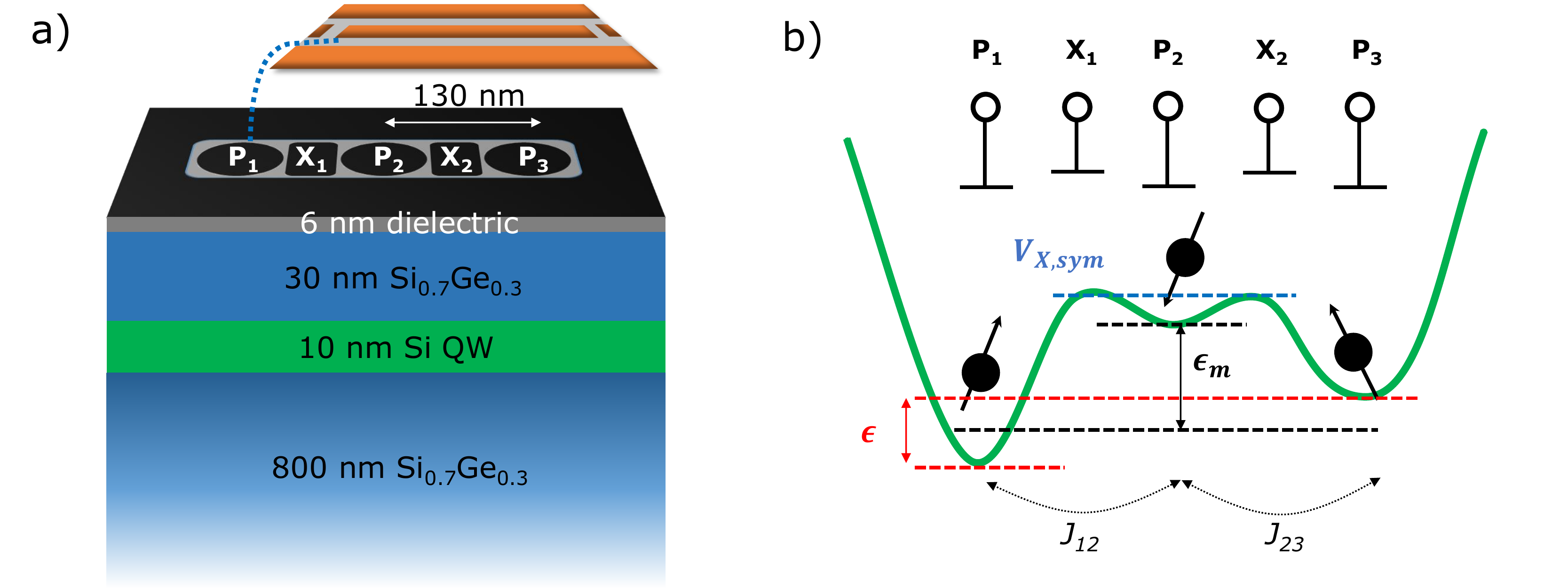}
	\centering
	\caption{(a) Schematic of a Si/SiGe TQD device with five control gates (three plunger P gates, two exchange X gates). For microwave coupling, we assume P1 is connected to a MW resonator. 
	Device dimensions listed here are used for FCI simulations described in the text.
	(b) Collective potential coordinates ``dimple'' $\epsilon_m$, edge detuning $\epsilon$, and exchange bias $\VXsym$ used for controlling the TQD in RX operation. 
	Exchange between electrons under P1/P2 and P2/P3 corresponds to the $J_{12}$ and $J_{23}$ exchange axes, respectively.
	}
	\label{fig:devgeom}
\end{figure}

We are interested in TQD devices like the structure depicted in Fig. \ref{fig:devgeom}(a), where electrons are confined vertically in a strained silicon quantum well (QW) surrounded by Si$_{0.7}$Ge$_{0.3}$ barrier regions. 
Qubit control is done using five gates: three ``plunger'' (P) gates to control the chemical potential of each dot and two ``exchange'' (X) gates to control the tunnel barriers between adjacent dots. 
A global screening gate surrounds the dot gates to isolate the quantum dot region but plays no other active role (we ignore design details such as overlapping gate geometries for simplicity \cite{borselli_undoped_2015,zajac_reconfigurable_2015}). 
In experimental devices, additional gates are also needed to control electron baths, neighboring charge sensors, etc.; we neglect these since their impact on exchange is usually minuscule. 
When evaluating spin-photon interactions, we will also assume that one dot gate---P1 unless otherwise noted---is connected to the voltage anti-node of a single-mode cavity, as commonly done in experiments \cite{mi_coherent_2018,samkharadze_strong_2018,landig_coherent_2018}; the voltage induced by a cavity photon perturbs the TQD through this gate, inducing electric coupling to the spin qubit.

The spin state of a three-electron TQD is described by three quantum numbers. 
In the most convenient basis, these are the total spin angular momentum $S$, the spin projection $m$ along a given axis (typically arbitrarily the $z$-axis), and the total spin $S_{ij}$ of electrons $i$ and $j$ in the TQD. 
The DFS encoded qubit is defined by the two lowest energy states in the $S = 1/2$ sector, with an extra gauge degree of freedom provided by $m = \pm 1/2$ \cite{divincenzo_universal_2000,fong_universal_2011}. 
Under decoherence-free subsystem operation, this gauge can be left unspecified without impacting exchange operation or photon coupling, assuming a completely homogeneous magnetic environment. 
For computational simplicity, our calculations are therefore performed assuming zero magnetic field and the arbitrary choice of $m = 1/2$.
Within this two-dimensional space, we can choose an electron pair whose collective spin $S_{ij} = {0,1}$ defines the qubit basis. The qubit can then be controlled by exchange and measured by Pauli spin blockade \cite{prance_single-shot_2012,jones_spin-blockade_2019}.
In practice the DFS qubit is controlled by manipulating gate voltages to tune $J_{12}$ and $J_{23}$; since there are five gates per qubit, this can be done in multiple ways. 
In exchange-only (EO) operation, one electron resides in each P dot and the different exchange pairs define two (nonorthogonal) qubit axes on the Bloch sphere, which can be sequentially modulated for full qubit control  \cite{eng_isotopically_2015,reed_reduced_2016}. 
By contrast, RX operation requires both exchange axes to be active simultaneously. 
In this mode the qubit is best controlled by the three collective bias coordinates shown in Fig. \ref{fig:devgeom}(b): the ``dimple'' detuning $\epsilon_m = V_{P2}-(V_{P1}+V_{P3})/2$, the detuning of the edge dots $\epsilon = V_{P1}-V_{P3}$, and the exchange bias $\VXsym = V_{X1} = V_{X2}$ (operating with symmetric tunnel couplings for simplicity). 
FH theory predicts favorable spin-photon coupling points lie at zero $\epsilon$ where the symmetry reduces charge noise dephasing \cite{russ_long_2015}. 
We will therefore focus on qubit control via the dimple and exchange bias coordinates. 

Each gate voltage configuration $\vec{V}$ corresponds to a device electrostatic Hamiltonian $\ham$ under which the TQD qubit evolves. 
The qubit energy splitting is set by the exchange energy
\begin{equation} \label{eq:Jdef}
hJ = \bra{1}\ham\ket{1} - \bra{0}\ham\ket{0},
\end{equation}
where $\ket{0,1}$ are the two lowest energy eigenstates. 
Other quantities of interest depend on the gate lever arm operators $H'_i = {d\ham}/{d V_i}$, which describe the Hamiltonian's response to small voltage perturbations on any gate $i$. 
In particular, the transverse spin-photon coupling rate $g$ is given by
\begin{equation}\label{eq:hg}
hg = \braket{1|H'_{P1}|0} V_0,
\end{equation}
where $V_0 = 2\pi f_r \sqrt{{\hbar Z_0}/{\pi}}$ is the root-mean-square voltage of a vacuum fluctuation in the fundamental mode of a cavity with characteristic impedance $Z_0$. Details about the derivation of this formula are given in \ref{gcalc}. Unless otherwise noted, we use $Z_0 = 50$ $\Omega$ as a baseline value for calculations. To maximize the coupling, we assume that the qubit is resonant with the cavity, \textit{i.e.}, the latter is chosen so the resonance frequency $f_r = J$. 

Qubit dephasing is another important consideration. 
We assume the dominant TQD dephasing channel for spin-photon coupling is electrical charge noise; magnetic noise (typically due to hyperfine gradients in Si/SiGe devices \cite{andrews_quantifying_2019}) also causes dephasing and leakage but is mostly suppressed at the high values of $J$ needed for spin-photon coupling.
In circuit QED, the qubit dephasing rate $\gamma$ is often extracted experimentally and interpreted as an effective Markovian decay parameter, which is valid for white noise channels but timescale- and experiment-dependent for colored noise. 
Empirically, semiconductor qubits are often limited by $1/f$ charge noise \cite{reed_reduced_2016,petit_spin_2018,connors_low-frequency_2019}, which in turn leads to a power spectral density of exchange fluctuations $S_J(f)=A_J^2 / f$. 
Therefore we will focus on modeling the expected exchange noise amplitude $A_J$; we discuss the connection between this quantity and $\gamma$ for different experiments in \ref{iSWAP}.
While the microscopic sources of charge noise remain under investigation, we can model it using gate-referred potential fluctuations \cite{reed_reduced_2016} under certain assumptions described in detail in \ref{gatenoise}. 
The final result parameterizes the noise in terms of derivatives of $J$ with respect to the TQD gates (including P and X gates) using
\begin{equation} \label{eq:AJ}
A_J^2 = A_{\mu}^2 \sum_{i}^{\Ngates} \frac{1}{\alpha_{i}^2} \left(\frac{\partial J}{\partial V_i}\right)^2,
\end{equation}
where $\alpha_i$ is the lever arm of each gate and $A_{\mu}$ is the magnitude of underlying $1/f$ fluctuations of the QW chemical potential.
In this paper we generally assume $A_{\mu} = 1$ ueV/$\sqrt{\textnormal{Hz}}$, which is typical of charge noise measurements in quantum dots  \cite{connors_low-frequency_2019,wu_two-axis_2014,freeman_comparison_2016}.

\subsection{Fermi-Hubbard Model of TQD Operation}
We base our FH Hamiltonian on the schematic TQD model of Fig. \ref{fig:devgeom}(b), with three electrons and nearest-neighbor tunnel coupling. 
The chemical potential in the QW under each gate $\mu_i$ is capacitively controlled by the voltages of all gates via
\begin{equation}
\mu_i=\sum_{j=1}^{\Ngates}\alpha_{i,j}V_j .
\end{equation}
The gate cross-capacitances, which have units of energy per voltage, are approximated as $\alpha_{i,j}=\alpha_{j} / 3^{|i-j|}$, a simple heuristic inferred from electrostatic simulations.
Each $\mu_{Pn}$ sets the energy of an electron under plunger $n$ while $\mu_{Xn}$ sets the barrier height for tunneling between dots $n$ and $n+1$. 
We use a WKB-like model to describe the tunnel coupling
\begin{equation}
t_{n,n+1}=t_0 \exp\left(-\frac{\mu_{Xn} - (\mu_{Pn} + \mu_{Pn+1})/2 - \delta\chi} {\chi_0}\right),
\end{equation}
with model parameters $t_0$, $\chi_0$, and $\delta\chi$. 
Finally, an on-site Coulomb repulsion $U_c$ describes the electrostatic energy penalty for two electrons to occupy the same dot. 

\begin{figure}
	\includegraphics[width=0.9\textwidth]{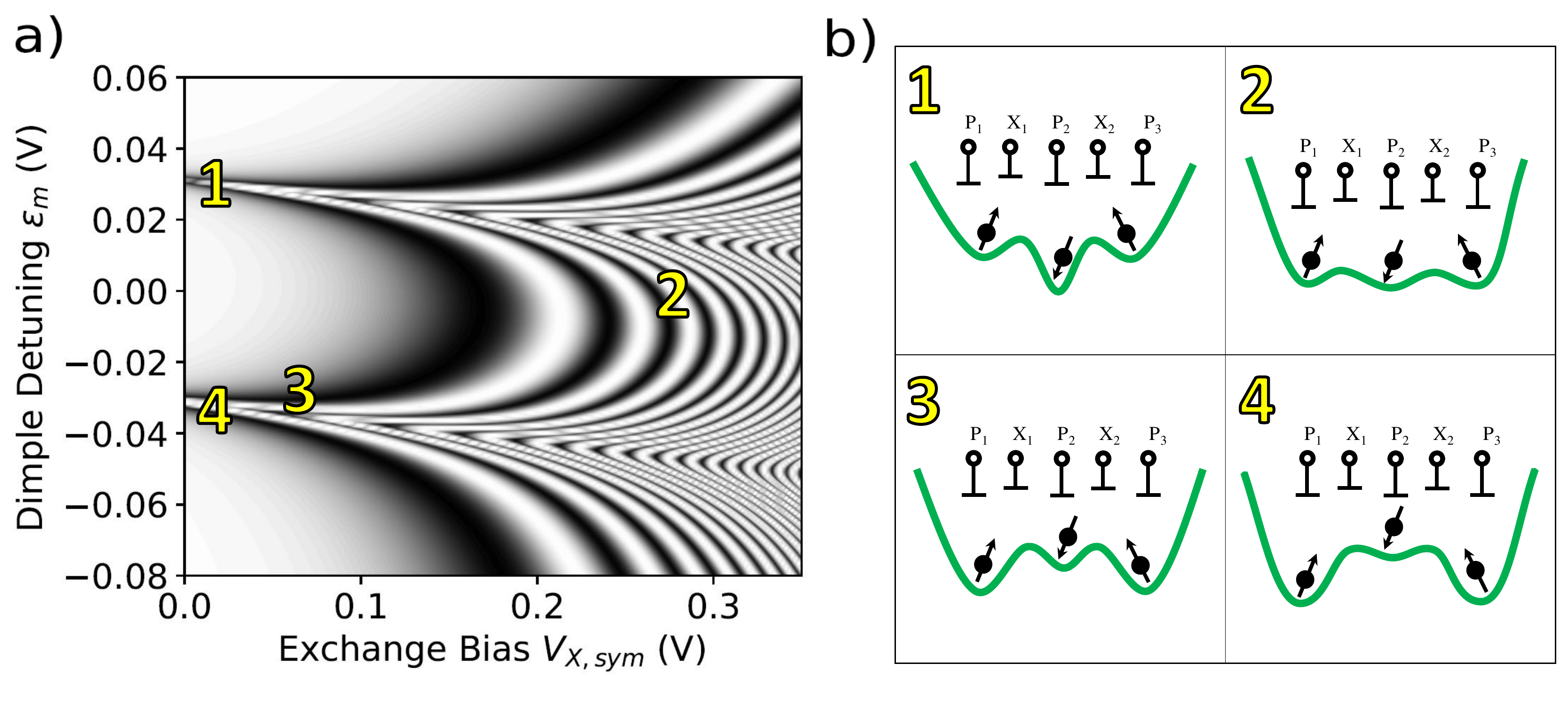}
	\centering
	\caption{(a) FH calculation of Rabi fringes of exchange ($= [1+\cos(2\pi Jt_{e})]/2$ for fixed evolve time $t_e = 1$ ns) versus dimple and exchange bias. The simulation parameters here are $\alpha_{Pn}=-0.1$ eV/V, $\alpha_{Xn}=-0.025$ eV/V, $t_0=10$ $\mu$eV, $\chi_0=2$ meV, $\delta\chi=3$ meV, and $U_c=4$ meV. Exchange increases rapidly as the dimple detuning approaches $\pm0.03$ V, leading to ``seams'' of fast Rabi oscillations where the (111) charge states anticross with the (120)-(021) and (201)-(102) charge states, respectively. (b) Illustrations of qualitative biasing conditions for different operating points.}
	\label{fig:FH_fingerprint}
\end{figure}

The Hilbert space of our FH Hamiltonian is spanned by eight states within the relevant spin subspace $S=m=1/2$.  Denoting the electron occupation in the P1, P2, and P3 dots as $\ket{ijk}$, we obtain two different spin states for the $\ket{111}$ charge configuration and six permutations of $\ket{012}$ dot occupancy. 
For RX operation, it is convenient to decompose the $\ket{111}$ states according to whether the ``edge'' spins 1 and 3 form a singlet or triplet, so $\ket{s}\equiv\ket{111,S_{13}=0}$ and $\ket{t}\equiv\ket{111,S_{13}=1}$. 
Pauli exclusion guarantees that the remaining low-energy states are spin singlets for the doubly-occupied dot. 
Note that for simplicity this model excludes valley and orbital excitations within each dot, which lead to higher-energy states where spin triplets share the same dot \cite{culcer_quantum_2010}. 
We also omit the $\ket{012}$ and $\ket{210}$ states, as these are not significantly populated in the low-$\epsilon$ regime explored here. 
In the resulting six-state basis, up to an overall energy, the FH Hamiltonian is
\begin{equation} \label{eq:FH_Hamiltonian}
\eqalign{H_{FH} =&
	(\mu_1-\mu_2)\left(\ket{201}\!\!\bra{201}-\ket{021}\!\!\bra{021}\right) \cr
	&+ (\mu_2-\mu_3)\left(\ket{120}\!\!\bra{120}-\ket{102}\!\!\bra{102}\right)\cr
	&+\frac{t_{12}}{2}\left(\ket{201}+\ket{021}\right)\left(\bra{s}+\sqrt{3}\bra{t}\right)+\textrm{h.c.}\cr
	&+\frac{t_{23}}{2}\left(\ket{120}+\ket{102}\right)\left(\bra{s}-\sqrt{3}\bra{t}\right) + \textrm{h.c.} \cr
	&-U_c(\ket{s}\!\!\bra{s} + \ket{t}\!\!\bra{t}).
}
\end{equation}
Magnetic field effects are ignored as a global field does not affect the states within a chosen gauge $m$. In practice a finite uniform magnetic field might be applied to break Zeeman degeneracy, which may be relevant for hyperfine or spin-orbit effects, but as the latter are not considered here, our results are field-independent.

\subsection{Simultaneous Exchange Operation in FH Models}
We can diagonalize Eq. \ref{eq:FH_Hamiltonian} to obtain $J$. 
In Fig. \ref{fig:FH_fingerprint}, we plot Rabi fringes of exchange computed in this way as a function of $\epsilon_m$ and $\VXsym$. 
This is essentially a plot of $\cos(2\pi Jt_e)$ for $t_e = 1$ ns, so each fringe is a multiple of 1 GHz. 
We visualize exchange in this way because it accentuates trends in the bias dependence and is related to how the qubit spectrum is often probed in practice \cite{reed_reduced_2016}.
We also indicate the schematic potentials corresponding to the key features observed for different exchange biases.
The RX region most often considered in theory and experiment is situated at zero edge detuning and negative dimple (such as state 3 in Fig. \ref{fig:FH_fingerprint}), where the (111) charge configuration is perturbatively admixed with excited (201) and (102) states. 
The ``seam'' of high exchange indicated by state 4 corresponds to the asymmetric RX (ARX) point \cite{russ_asymmetric_2015} where these three charge states anti-cross.
Analogous behavior occurs at the opposite charge boundary denoted by state 1 due to admixing of the (111)-(120)-(021) states.
Finally, state 2 lies in the AEON (always-on exchange) mode in the middle of the (111) charge cell \cite{shim_charge-noise-insensitive_2016}.

\begin{figure}
	\includegraphics[width=0.9\textwidth]{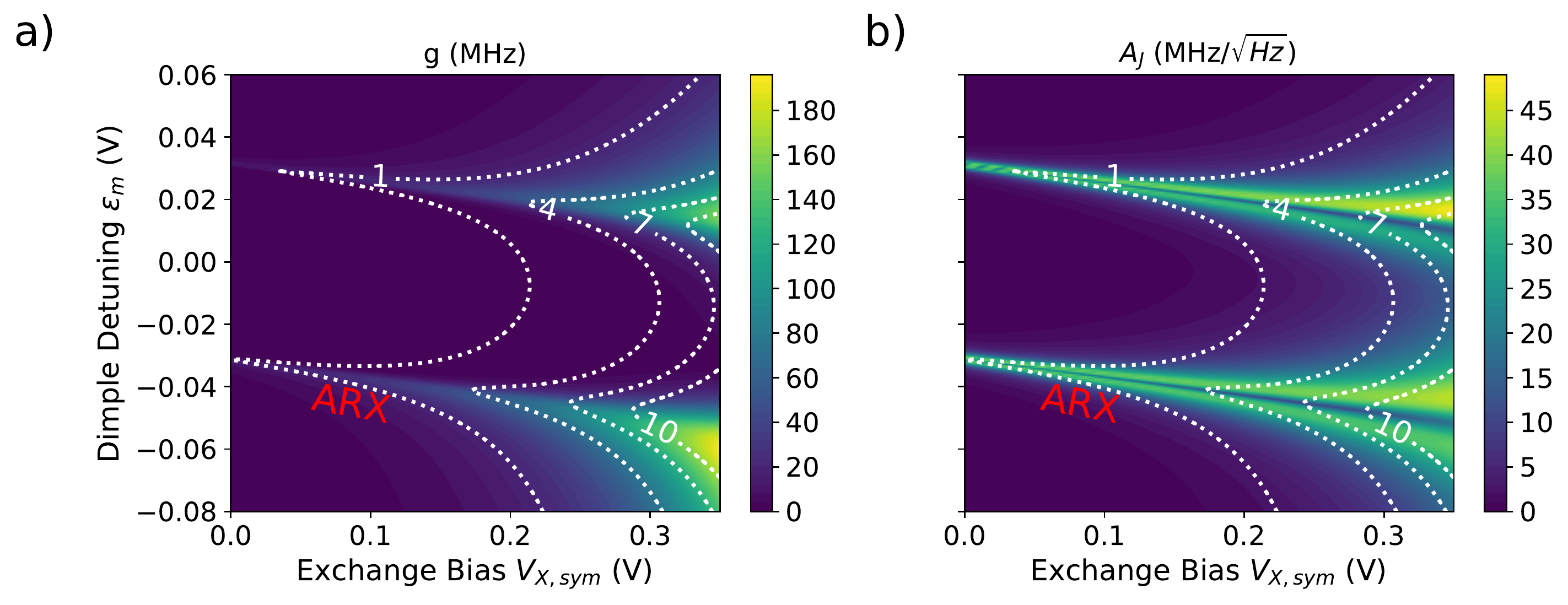}
	\centering
	\caption{(a) Spin-photon coupling rate $g$ and (b) $1/f$ $J$ noise amplitude $A_J$ from FH calculations. Contours of constant $J$ between 1-10 GHz are indicated by dashed white lines. $A_{\mu} = 1$ ueV/$\sqrt{\textnormal{Hz}}$ is assumed for noise calculations. The two narrow features with finite $g$ and minimum $A_J$ at positive and negative dimple occur at the (1,1,1) charge boundaries; we highlight the latter case as the main ARX regime of interest.}
	\label{fig:FH_coupling}
\end{figure}

Since $H_{FH}$ can be quickly diagonalized, it is easy to numerically compute its voltage derivatives to find the spin-photon coupling rate $g$ and exchange noise amplitude $A_J$, shown in Fig. \ref{fig:FH_coupling}.
We are typically interested in operating where $g$ is large and $A_J$ is minimal, as these allow us to achieve the most photon coupling before decohering. 
Narrow regions fulfilling these conditions indeed occur at the ARX points, as previously predicted theoretically  \cite{russ_long_2015,srinivasa_entangling_2016,russ_coupling_2016}.
For instance, along the ARX seam at negative dimple, the admixture of (111), (201), and (102) in the qubit states allows for interdot charge transitions when the state is perturbed by an electric dipole, leading to a large transverse photon coupling.
The symmetry of this hybridized charge state simultaneously offers a degree of protection against noise, leading to a local dephasing sweet spot.
The FH model also predicts similar behavior at positive dimple when the (111), (120), and (021) states admix \cite{russ_three-electron_2017}; we will mostly focus on the negative dimple ARX transition, expecting that under realistic biasing conditions, the potential here maximizes the electron separation and dipole coupling.
A major goal of our FCI calculations in the next section is to see whether and how these features change under detailed simulation.

\section{\label{operation}Device Modeling of TQD Operation with FCI}
\subsection{FCI Device Simulation Methodology}
The parameters of FH models are physically meaningful but sensitive functions of the device design and tuneup, making them difficult to quantitatively predict from first principles. 
By contrast, we can calculate TQD exchange exactly by solving the full 3-D three-electron Hamiltonian
\begin{equation}\label{eq:fciham}
\eqalign{
	\ham &= \sum\limits_{m=1}^3 \left(-\frac{\hbar^2}{2 m_x}\frac{ \partial^2}{\partial x_m^2} -\frac{\hbar^2}{2 m_y}\frac{ \partial^2}{\partial y_m^2} -\frac{\hbar^2}{2 m_z}\frac{ \partial^2}{\partial z_m^2} + e\phi_{\vec{V}}(\vec{r}_m) \right) \cr
	&+ \frac{1}{2}\sum\limits_{m \neq n} \frac{e^2}{4\pi \epsilon_r |\vec{r}_m - \vec{r}_n|}
}
\end{equation}
where the single-particle potential $\phi_{\vec{V}}(\vec{r})$ includes both the heterostructure confinement and the electrostatic potential induced by the gate voltages $\vec{V}$. This can be generalized to include image effects and other self-consistent electrostatic interactions, though we neglect them as minor perturbations for the effects we consider here.

To solve Eq. \ref{eq:fciham} we use a device simulator based on the FCI method, a powerful technique originally developed for quantum chemistry problems \cite{szabo_modern_1996} and well-suited for analyzing few-electron quantum dots \cite{nielsen_configuration_2010,nielsen_implications_2010,kim_optically_2016}. 
Since the RX operating regime generally lies within relatively narrow voltage ranges, dense and accurate sweeps over bias space are required to resolve the features of interest. 
Our FCI code is tailored to efficiently compute exchange in planar quantum dots and able to accurately resolve energy splittings down to the neV scale; details about the computational scheme can be found in \cite{c._anderson_et_al._high_nodate}. 
Equally important is the incorporation of 3-D device electrostatics rather than analytic potentials to describe real device operation.
For each set of gate biases $\vec{V}$, we solve the Poisson equation to obtain the electrostatic potential $\phi_{\vec{V}}(\vec{r})$, which is used to compute the electronic spectrum in FCI. 
Unless otherwise noted, all FCI calculations in this paper are based on the device structure depicted in Fig. \ref{fig:devgeom}(a). 

Since we are focused on how electrostatics impact exchange, we ignore valleys and model each electron with a single anisotropic effective mass Hamiltonian (valley physics can be incorporated with appropriate extensions of this Hamiltonian \cite{culcer_quantum_2010}). 
This effectively assumes a device with large and uniform valley splitting; the impact of small and/or spatially nonuniform valley splittings is certainly important and would further complicate silicon qubit operation, but lies outside the scope of this work. 
As before, we ignore magnetic fields and we obtain $J$ from the lowest energy eigenstates of the $m = 1/2$ spin sector. 
We can also directly evaluate the transverse spin-photon coupling rate $g$ from the microscopic three-electron wave functions computed by FCI using Eq. \ref{eq:hg}. 
This requires the spatial lever arm operator for the cavity-coupled gate $H'_{P1}(\vec{r}) = e{\partial \phi_{\vec{V}}(\vec{r})}/{\partial V_{P1}}$. 
Since $H'_{P1}$ is a single-electron operator and $\ket{0, 1}$ are three-electron eigenstates, the matrix element $\braket{1|H'_{P1}|0}$ required for $g$ can be computed using standard methods \cite{szabo_modern_1996}.

The effects of charge noise can be similarly estimated from the qubit states using Eq. \ref{eq:AJ}, which requires the gate voltage derivatives of $J$ to define noise sensitivity. 
From first-order perturbation theory, the derivative of $J$ with respect to the voltage of a particular gate $i$ is
\begin{equation}
\frac{\partial J}{\partial V_i} = \braket{1|H'_i(\vec{r})|1} - \braket{0|H'_i(\vec{r})|0} = \int \delta n_J(\vec{r}) H'_i(\vec{r}) d\vec{r},
\end{equation}
where $\delta n_J(\vec{r}) = n_1(\vec{r}) - n_0(\vec{r})$ is the differential spatial charge density of the qubit states at a given voltage configuration. 
Since the lever arm operators can be precomputed, this significantly reduces the computational load by allowing us to obtain all relevant derivatives of $J$ from a single FCI calculation. 
The differential density also allows us to visualize the charge multipoles of exchange for further intuition into noise susceptibility, as discussed in \ref{multipoles}.

\begin{figure}
	\includegraphics[width=1\textwidth]{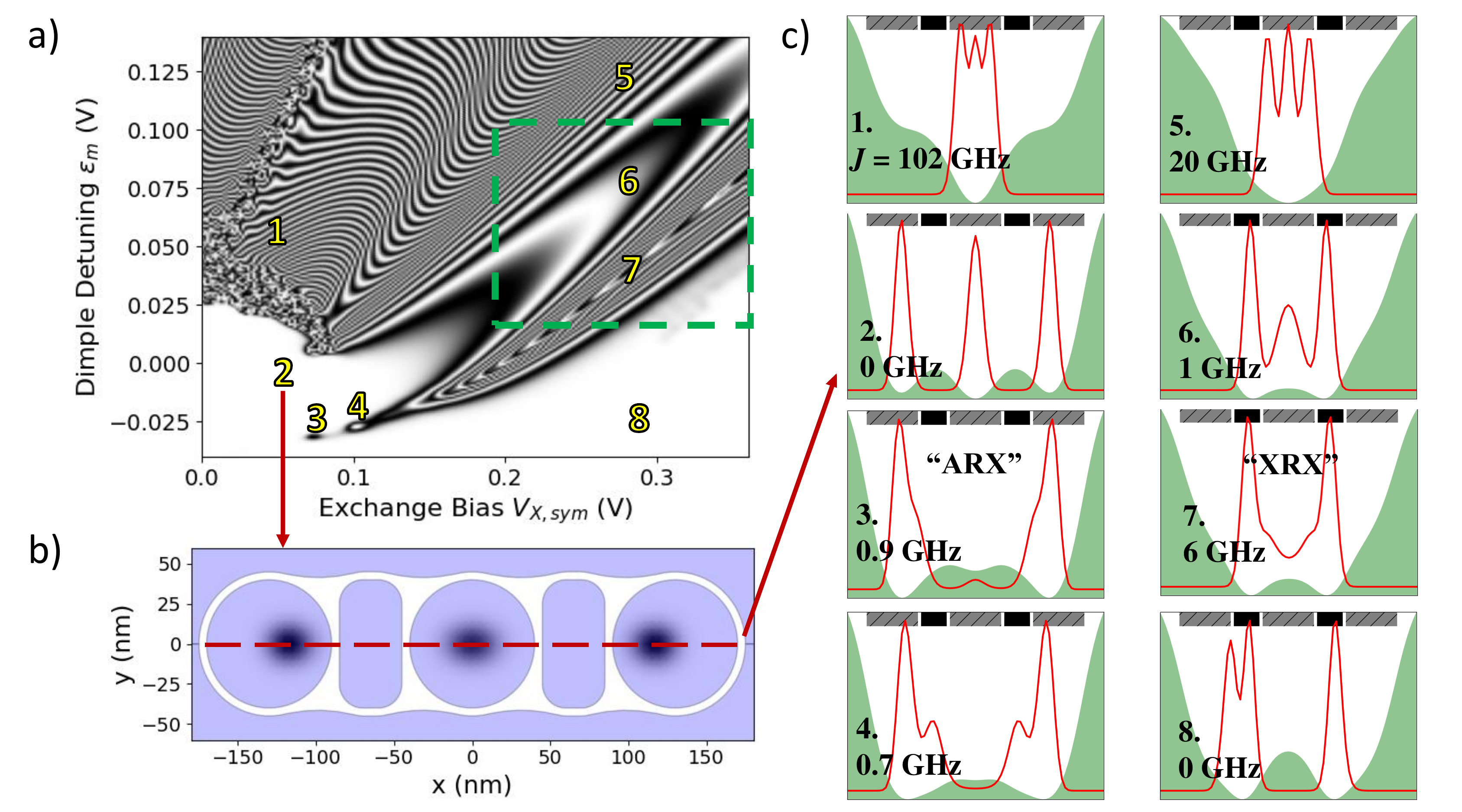}
	\centering
	\caption{a) Si/SiGe TQD Rabi fringes ($= [1+\cos(2\pi Jt_{e})]/2$ for fixed evolve time $t_e = 1$ ns) versus dimple and exchange bias computed by FCI. 
	The dashed box indicates region where $g$ and $A_J$ calculations are done in Fig. \ref{fig:fcigAj}. 
	b) ``Top view'' of FCI electron density in the QW region at bias point 2 with gate geometry superimposed in blue. 
	Cuts along the (dashed horizontal) center axis can be used to extract c) the associated device potential (green) and ground state density (red) at the labeled bias points in (a), along with the computed $J$ for these points. 
	The filled (hatched) boxes at the top of each cut mark the positions of the X (P) gates above the well. 
	Biases 3 and 7 are examples of the ARX and XRX states which are our primary focus.}
	\label{fig:fcifinger}
\end{figure}

\subsection{Simultaneous Exchange Operation in Si/SiGe TQDs from FCI}
In Fig. \ref{fig:fcifinger}(a), we show the behavior of the Rabi fringes of exchange calculated by FCI for a silicon TQD; again, each fringe corresponds to a multiple of 1 GHz. 
We observe qualitative differences from the Fermi-Hubbard results in Fig. \ref{fig:FH_fingerprint} due to the malleability of electron position in realistic gate-defined potentials.
To visualize these effects, the electrostatic potential and ground state electron density at various bias points are depicted in Fig. \ref{fig:fcifinger}(b)-(c); for reference, the lateral P and X gate positions are marked with hatched and solid boxes in each graph. 
At high positive dimple bias and low $\VXsym$, such as state 1 denoted in Fig. \ref{fig:fcifinger}(c), we observe that electrons congregate in the middle plunger dot P2, leading to very large $J$ and finely spaced fringes. 
By contrast, exchange is suppressed when electrons are evenly distributed along the TQD even when $\VXsym$ is ramped to increase tunnel coupling, as indicated by the lack of Rabi fringes in the vicinity of state 2 in Fig. \ref{fig:fcifinger}(a). 
The very sharp transition with positive dimple from zero to high exchange between states 2 and 1 when the qubit tips over from (111) to multiple electrons in the center dot makes it hard to control the ARX-like crossing at the charge boundary. 
These observations are key to understanding our subsequent findings since they imply the difficulty of tuning $J$ in typical silicon devices while keeping electrons localized under neighboring plunger gates; the relatively high effective mass of silicon ($m^*_t = 0.2$) suppresses kinetic exchange between electrons separated at distances of 100 nm and above. 
As our calculations will show, barrier modulation (\textit{e.g.}, forward biasing an X gate) increases $J$ in silicon dots primarily by softening the potential to reduce the electron separation, enhancing their Coulomb interaction, rather than by simply increasing the tunnel coupling through a fixed barrier, as might be intuited from simple FH models. 
Our observations in simulations are consistent with recent experiments showing the importance of electron displacement in modulating simultaneous exchange in GaAs \cite{qiao_coherent_2020}; as discussed in Section \ref{spinphoton}, such effects are even stronger in silicon owing to its larger mass.

Biasing closer to the RX operating regime (more negative dimple and higher exchange bias) leads to a subtle feature in the FCI Rabi plot, corresponding to state 3 in Fig. \ref{fig:fcifinger}(c). 
This state with $J = 1$ GHz looks like the expected ARX configuration, which is a (201)-(111)-(102) hybridized state with most of the electron density residing under the edge plungers but some charge shared under P2. 
Moreover, we compute a sizable dipole moment $\braket{1|x|0} = 57$ nm between the eigenstates at this bias, indicating the potential for strong transverse coupling. 
However, in contrast to FH calculations where dense Rabi fringes around the RX region imply robust exchange, this state is isolated in bias space. 
Fine adjustments in device tuneup around this point tend to tip the qubit into charge configurations like state 2 (balanced TQD) or state 4 (electrons totally localized under P1 and P3), either of which suppresses $J$ due to the electron spatial separation. 
State 4 occurs for instance when we seek to increase $J$ of the ARX state by forward biasing $\VXsym$; the X gate cross capacitances instead smooth out and remove the quantum dot confinement under P2. 
The low energy eigenstates are then nearly degenerate combinations of (201) and (102) which are only weakly coupled through the barrier formed by P2. 
This underscores the fragility of ARX operation in silicon, where the weak electron overlap leads to a cutoff in accessible $J$ of about 1~GHz for the design considered here. 
This cutoff can be increased somewhat by reducing the gate pitch to force electrons closer together and enhance exchange. 
However, such changes cause tradeoffs in device operation by lowering gate lever arms and increasing cross-capacitances. 

Turning to TQD operation at even higher $\VXsym$, new features in Rabi evolution become evident in Fig. \ref{fig:fcifinger}(a). 
At large $\epsilon_m$, very high $J$ values in the tens of GHz occur as all three electrons are forced under the middle plunger (state 5). 
Reducing the dimple distributes the electrons more evenly underneath the X gates and P2 (state 6). 
The device enters a AEON-like operating regime here, similar to the pronounced lobe of FH Rabi fringes in Fig. \ref{fig:FH_fingerprint}.
In contrast to the conventional TQD AEON point \cite{shim_charge-noise-insensitive_2016}, however, the effective (111) charge state here actually corresponds to electron positions under the barrier rather than edge plunger gates. 
The greater spatial proximity of the electrons robustly allows for $J > 1$ GHz, in contrast to the conventional TQD (111) tuneup (like state 2) where exchange is suppressed. 

As we continue to reduce the dimple bias, another seam of high exchange forms where $J$ ranges between 1-10 GHz within the plotted range. 
In this region, corresponding to features like state 7 in Fig. \ref{fig:fcifinger}(c), the electrons reside within an ``inverted'' DQD underneath the barrier gates. 
The exchange or qubit splitting here is directly set by the DQD tunnel coupling due to the effective potential barrier controlled by P2.
Intuitively, two of the electrons in the XRX state are bound on either side of the DQD while the third electron tunnels between the dots, undergoing exchange with both. 
We will denote this state as ``XRX'' because the electrons behave similarly to the RX regime but are now localized under the X gates. 
Importantly, since XRX resembles a three-electron charge qubit, we find that this state also has a finite transverse dipole matrix element $\braket{1|x|0} = 27$ nm, allowing it to couple to MW photons. 
In principle we can access the hybrid qubit regime by detuning the X gates asymmetrically to tip the third electron into either side \cite{shi_fast_2012}, though we do not explore this as the transverse coupling is largest at zero detuning. 
The cross capacitances of the X and P2 gates cause exchange to rise when $\epsilon_m$ and $\VXsym$ are increased, as the voltage on P2 (which scales with $\epsilon_m$) mostly modulates the tunnel coupling. 
Finally, as the dimple bias is further decreased, tunneling is suppressed between localized electrons under either X gate, turning off exchange. 
In principle, the ground state at precisely zero detuning should remain a linear combination of (201) and (102) where the edge dots lie under X1/X2, though the tunnel coupling is so small in practice that even minute numerical errors may lead to asymmetric states in calculations, as seen for state 8 of Fig. \ref{fig:fcifinger}(c). 
We note that such regions of suppressed evolution are unlikely to be observed experimentally, since any finite detuning would deposit the qubit in either the ground (201) or (102) singlets, which undergo fast exchange with excited triplets within the same charge state. 

While quantitative values will vary with device geometry, we expect our qualitative observations of a maximum cutoff $J$ for ARX and the existence of the XRX three-electron state to be common features in silicon devices. 
Overall, the rich behavior found in Fig. \ref{fig:fcifinger} underscores the importance of realistic device electrostatics for understanding exchange physics missed by simpler models. 
In particular, the role of electron proximity in modulating $J$ in silicon significantly changes the nature of simultaneous exchange regimes such as RX, ARX, and AEON, which may modify operating strategies based on these features (\textit{e.g.}, \cite{russ_three-electron_2017,shim_charge-noise-insensitive_2016,ruskov_quantum-limited_2019}). 
Such observations are difficult to predict in phenomenological models without detailed knowledge of all possibly relevant charge states and their bias dependence. 
Of course, our calculations remain very idealized since they do not account for device disorder and valley mixing, both of which are relevant in practice and could potentially obscure or destroy some of the subtler effects observed here. 
As a simple example of how asymmetries affect exchange, we discuss the impact of edge detuning $\epsilon$ in simulation in \ref{detuning}. 
Overall we expect the existence of the XRX state to be more robust than the ARX state against small perturbations, since it essentially occupies a distinct charge cell.
This also potentially enables easier experimental identification of this state starting from equilibrium charge stability cells.

\section{\label{spinphoton}Spin-Photon Coupling in RX Operation Modes}
\subsection{Computing $g$ and $A_J$ in Silicon and GaAs TQDs}
\begin{figure}
	\includegraphics[width=0.9\textwidth]{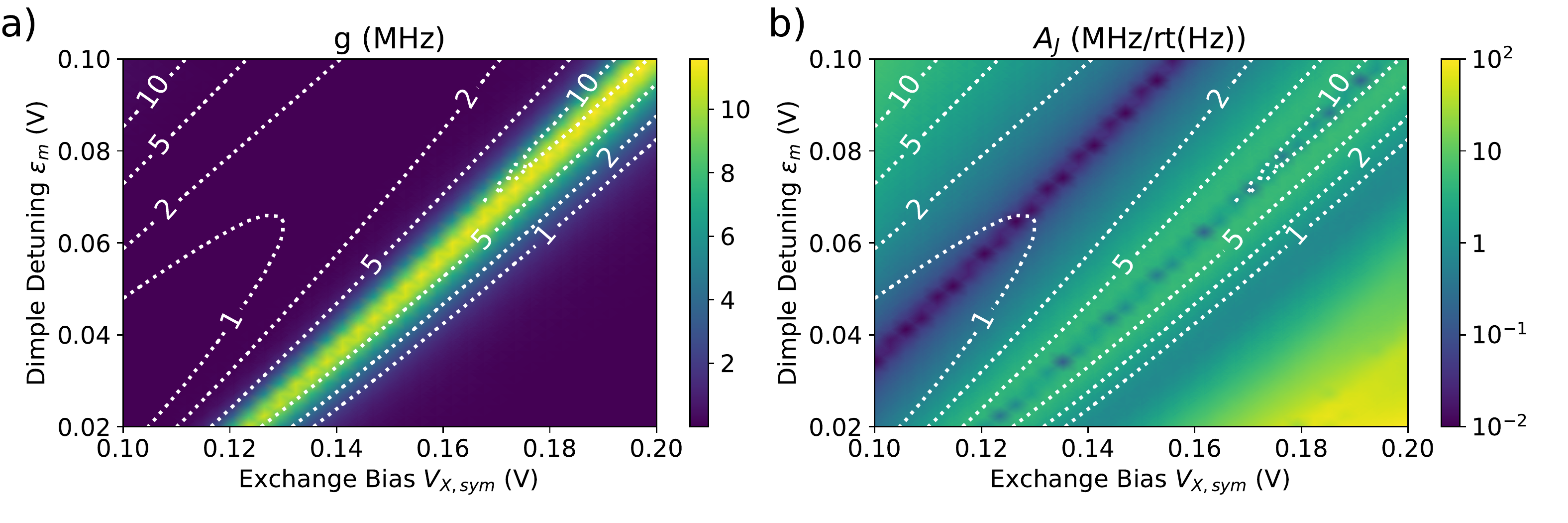}
	\centering
	\caption{FCI calculations of (a) spin-photon coupling rate $g$ and (b) $1/f$ charge noise amplitude $A_J$ as a function of dimple and exchange bias within the boxed region of Fig. \ref{fig:fcifinger}. 
	Iso-contours of exchange $J$ (in GHz) are overlaid for reference. 
	The value of $g$ at each bias assumes dot-photon coupling via a plunger gate to a 50 $\Omega$ resonator which is frequency matched with the qubit $J$. 
	Noise calculations assume fluctuation amplitude $A_{\mu} = 1$ ueV/$\sqrt{\textnormal{Hz}}$.}
	\label{fig:fcigAj}
\end{figure}
Our calculations show that, for the type of device design considered here, multi-GHz RX operation in silicon TQDs is difficult to obtain in conventional charge configurations and may instead mostly occur in XRX. 
To determine how this impacts spin-photon coupling, in Fig. \ref{fig:fcigAj} we compute $g$ and $A_J$ using the FCI eigenstates. 
Along the XRX seam, we observe $J \approx 5$-10 GHz within the simulated voltage range, allowing matching to typical MW resonator frequencies $f_r$. The corresponding magnitude of $g$ at these points varies over roughly 5-12 MHz. 
This is reasonable since, as noted before, the XRX state is essentially a three-electron charge qubit with a corresponding charge dipole moment (albeit one smaller than that of an ARX state because of the narrower electron spread).

Turning to calculations of $1/f$ exchange noise $A_J$, we observe a more complex bias dependence. 
Within the AEON-like regime running along the upper diagonal of Fig. \ref{fig:fcigAj}(a)-(b), the uniform charge distribution suppresses charge noise but also the photon coupling, leading to a distinct sweet spot with zero first-order transverse coupling. 
Dephasing increases as the dimple bias is reduced at fixed $\VXsym$, moving towards the charge-qubit-like XRX regime. 
Notably, there are signs of another (very) narrow charge noise ``sweet spot'' along the center of the XRX seam, the resolution of which is limited by the bias discretization in simulation (about 2 mV in $\epsilon_m$ in this figure). 
In \ref{multipoles} we discuss how this subtle feature is due to cancellation of the DC quadrupole moment owing to the increased screening of three-electron states, in contrast to one-electron DQDs where sweet spot operation at zero detuning only cancels the dipole moment. 
Given the fineness of this feature, higher-order noise effects not considered here may also be non-negligible in this regime \cite{russ_asymmetric_2015,russ_coupling_2016}.
Eventually the dephasing rate drops in concert with exchange as negative dimple detuning suppresses tunneling. 
The patch of fast dephasing in the lower right corner of Fig. \ref{fig:fcigAj}(b) is an artifact as it occurs at very low values of $J$ below the resolution of our calculations, where small numerical detunings lead to large differences in the ground and excited state densities.

While we focus on the logical states in our discussion, excited states are also relevant to device operation. 
Our calculations have ignored valleys for simplicity. 
In practice, high valley splittings are clearly desirable (at least significantly larger than the electron temperature and the highest $J$ of interest--here about 40 ueV) to avoid thermal excitations during operation. 
Maximizing valley mixing uniformity is also important since the relevant states will sample the full triple-dot region. 
For sufficiently large homogeneous valley splittings, orbital states may be another limiter. 
At the XRX points simulated here, the energy gap from ground to the lowest $S=1/2$ orbital excitation ranges between 150-250 ueV; however, since multi-electron orbital energies are sensitive to electrostatic confinement, these excitations can be tune-up-dependent in practice. 
These considerations also imply that the bias ramp between DFS and XRX should be optimized to avoid inducing valley-orbit transitions.

\begin{figure}
	\includegraphics[width=0.9\textwidth]{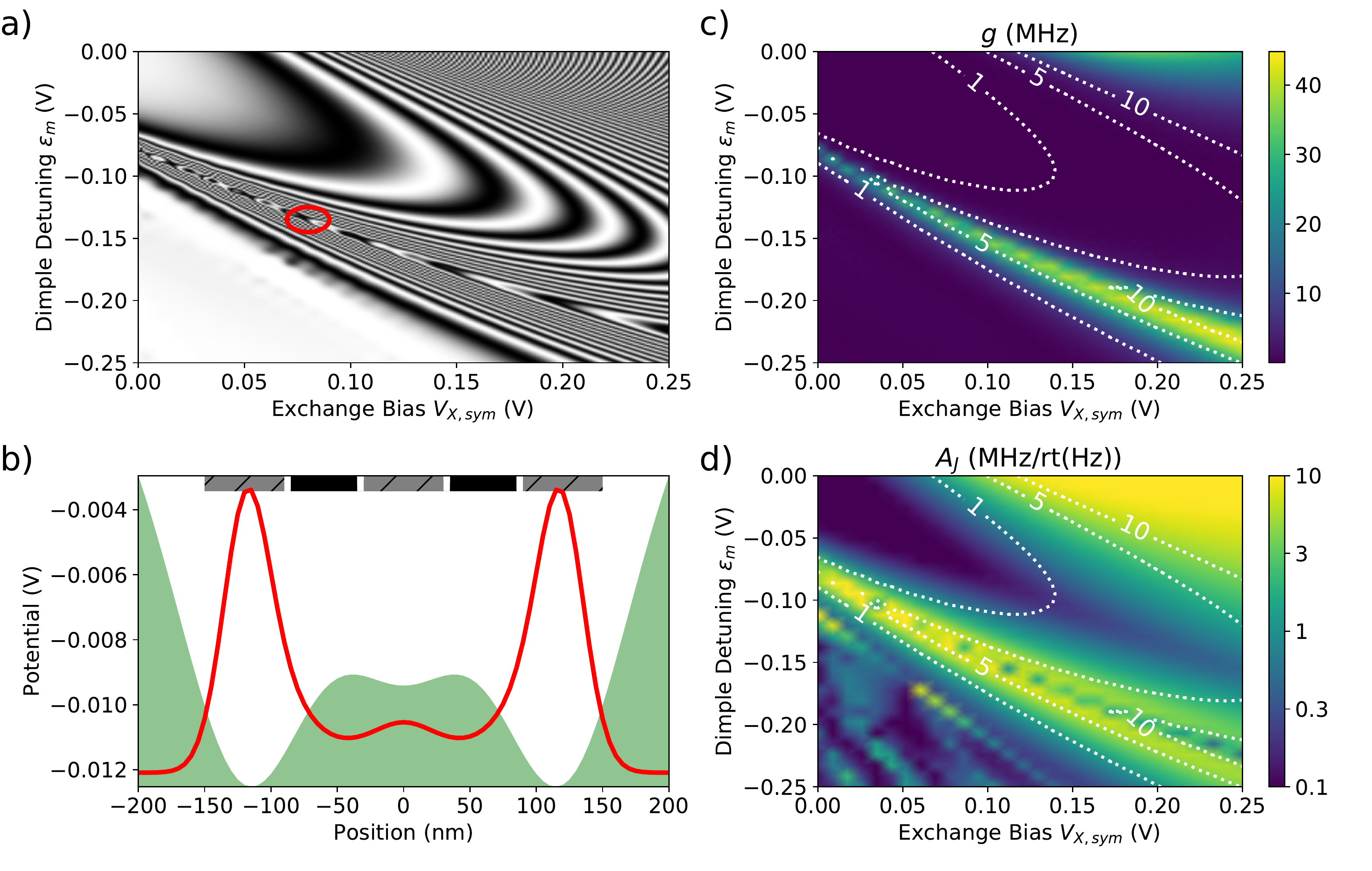}
	\centering
	\caption{FCI TQD calculations for GaAs TQD with approximate dimensions from \cite{landig_coherent_2018}. 
	(a) Rabi fringes of computed exchange versus $\epsilon_m$ and $\VXsym$. 
	Red circle indicates the bias at which we extract (b) the calculated electrostatic potential (green) and overlaid electron density (red) for $J = 4.5$ GHz. 
	Note this state lies along an ARX ``strip'' that extends to higher $J$ with increasing $\VXsym$. 
	(c) Spin-photon coupling rate $g$ and (d) $1/f$ exchange noise amplitude computed assuming the RF plunger gate is connected to a $Z_0 = 1.3$ k$\Omega$ resonator and $A_{\mu} = 1$ ueV/$\sqrt{\textnormal{Hz}}$. Contours of $J$ between 1-10 GHz exchange are overlaid in white. Spurious peaks in the lower left of this plot are due to numerical errors in the zero-$J$ limit.}
	\label{fig:fciGaAs}
\end{figure}

Having seen how exchange depends on the particular properties of silicon TQDs, it is useful to investigate how the choice of device material impacts the results. 
This is especially so since experiments in GaAs TQDs have already shown strong spin-photon coupling following FH-based prescriptions for accessing the RX state \cite{landig_coherent_2018,landig_virtual-photon-mediated_2019}.
Intuitively we expect kinetic exchange to be stronger in GaAs owing to its smaller in-plane effective mass compared to that of strained Si (0.067 vs. 0.2), allowing larger $J$ to be reached at greater electron separations. 
To assess the impact of this effect, we repeat our FCI simulations for a GaAs TQD with dimensions commensurate to the device depicted in Ref. \cite{landig_coherent_2018}. 
Interestingly, the resulting Rabi fringes in Fig. \ref{fig:fciGaAs}(a) appear qualitatively very similar to those of the FH calculations in Fig. \ref{fig:FH_fingerprint}, with an ARX ``seam'' visible below the main lobe of Rabi fringes. 
As FH predicts, this seam corresponds to a region of large spin-photon coupling seen in Fig. \ref{fig:fciGaAs}(c). 
From our simulations we extract a calculated value of $g =$ 27 MHz at $J =$ 4.5 GHz (coupled to a 1.3 k$\Omega$ resonator as used in experiment), in good agreement with the measured values of $g = 23-31$ MHz at 4.38 GHz \cite{landig_coherent_2018}. 
The indistinct signs of a charge noise sweet spot within this region are also consistent with FH calculations as well as the experimental observations (which are also complicated by strong hyperfine dephasing in GaAs). 
Finally, we observe in Fig. \ref{fig:fciGaAs}(b) that the calculated device potential and electron density at this point resemble those expected for ARX. 
These results show how device and material details affect the validity of FH models, and in particular how the lighter effective mass of GaAs makes conventional RX operation easier by enhancing kinetic exchange.

\subsection{Projecting Entanglement Protocol Performance in XRX}
Returning to silicon, the XRX regime found by our numerical modeling can be used for spin-photon coupling by ramping a qubit prepared in the conventional TQD DFS into the XRX state.
Given the distinct electron configuration in XRX compared to the typical TQD, one may ask how these states are related to the DFS basis, as is needed for spin transduction. 
For small voltage throws in EO operation where the electrons in the TQD remain well-separated, the lowest energy eigenstates are effectively orthogonal spin states and hence can be used as a logical basis. 
By contrast, the eigenstates under large simultaneous voltage throws, such as those used to reach XRX, exhibit more complex spin-charge entanglement. 
However, because these eigenstates at high voltage are adiabatically connected back to spin states in the well-separated TQD limit, we can still effectively identify them with logical spin states.

\begin{figure}
	\includegraphics[width=0.9\textwidth]{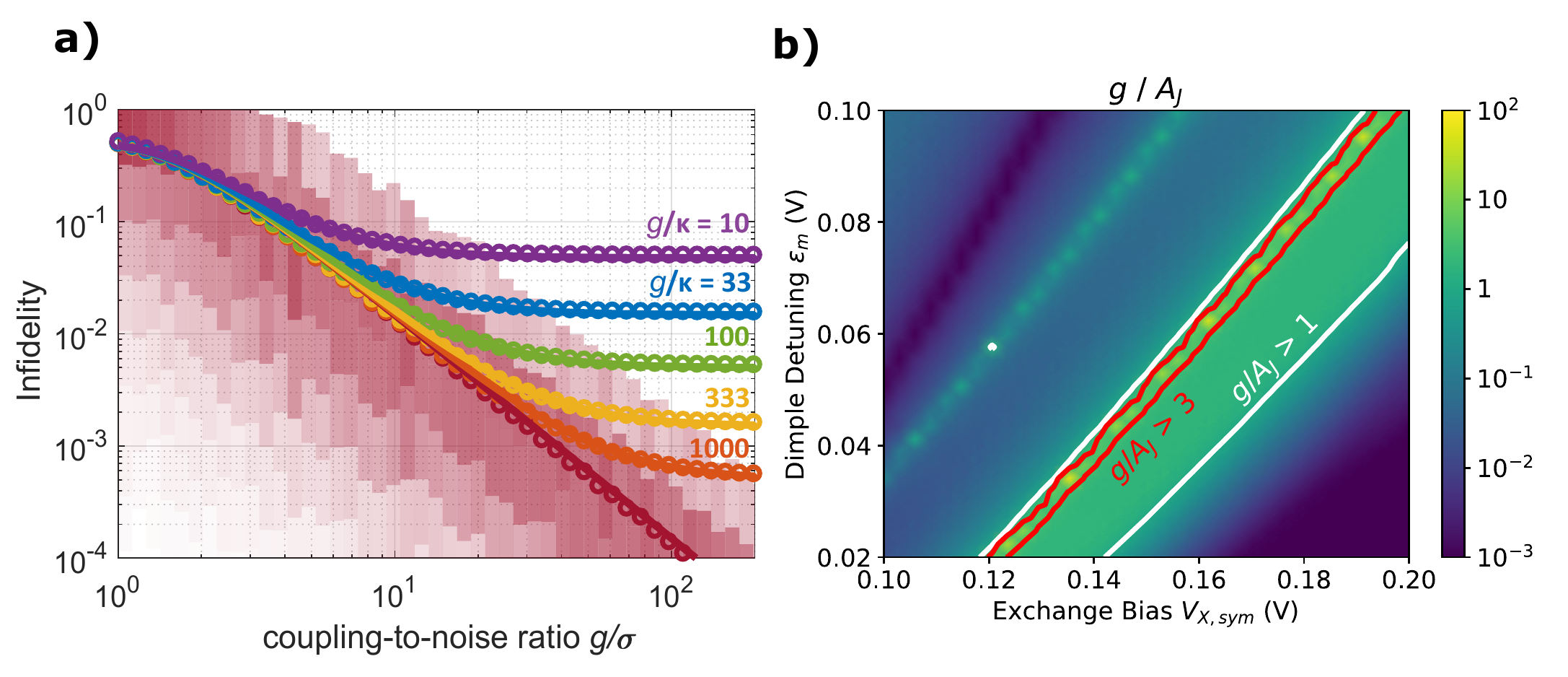}
	\centering
	\caption{(a) Simulated Bell-state infidelity $1 - \textrm{Tr}_{\textrm{\tiny cavity}}[\bra{\Psi^-}\rho\ket{\Psi^-}]$ following cavity iSWAP protocol as a function of the ratios of coupling-to-cavity-loss $g / \kappa$ and coupling coupling-to-detuning deviation $g/\sigma$ (see \ref{iSWAP} for more discussion). 
	$\sigma$ is related to the $1/f$ exchange noise amplitude $A_J$ via the low-frequency cut-off $f_L$ as $\sigma=A_J \sqrt{\log(g/f_L)}$. 
	Calculations are made for an ensemble of normal-distributed detuning values; the associated infidelity histogram for the dark red case, with $\kappa=0$, is shown as shaded rectangles. 
	The circles show the average of this distribution for each $ g/\kappa$. 
	The solid lines show the analytic model of Eq.~\ref{cqed_analytic}, which closely tracks the numerical results.
	(b) $g / A_J$ ratio extracted from Si/SiGe TQD FCI calculations of Fig. \ref{fig:fcigAj} with regions with values greater than 1 and 3 denoted.
}
	\label{fig:iswap}
\end{figure}

To evaluate the effectiveness of the calculated coupling, we consider a cavity iSWAP protocol for entangling distant spin qubits placed at separate anti-nodes of a superconducting resonator. 
This protocol consists of initializing qubit A into its excited state, biasing to activate the Jaynes-Cummings iSWAP interaction for time $1/8g$ (implementing a rt-iSWAP between qubit $A$ and a cavity photon), and then deactivating qubit $A$ and activating qubit $B$ for time $1/4g$ (realizing a iSWAP between the cavity and qubit $B$ starting from the ground state of the latter). 
We assume the dominant noise is low-frequency fluctuations of exchange $J$, resulting in fluctuation of the qubit detuning relative to the fixed cavity resonance $f_r$. 
The simulation procedure for this protocol in the presence of $1/f$ noise is described in \ref{iSWAP} and its results are shown in Fig.~\ref{fig:iswap}(a). 
If the detuning fluctuation variance is $\sigma^2$, then the fidelity of a Bell state resulting from this protocol can be roughly derived as
\begin{equation}
F\approx \frac{1+e^{-\kappa/g-3(\sigma/g)^2}}{2},
\label{cqed_analytic}
\end{equation}
where $\kappa$ is the cavity loss rate in Hz units, related to the cavity $Q$ as $\kappa=f_r/Q$. This simple formula agrees well with the simulations described in \ref{iSWAP}, as shown in the figure.
Simulations indicate, not surprisingly, that $1/f$ noise and quasistatic noise (in which qubit-cavity detunings are random but constant in time) behave similarly, since the $1/f$ case is dominated by the low-frequency part of the noise spectrum. 
The effective variance is therefore integrated as $\sigma^2 \approx A_J^2\log(g/f_L)$ where $f_L$ is a low-frequency cut-off corresponding to the inverse of a total averaging time in an experiment. 
The relationship between these quantities and commonly measured dephasing rates such as spectroscopic linewidths is discussed in the appendix.
Overall, these simulations indicate that avenues for improvement exist if shaped detuning pulses, which compensate for low-frequency detuning drift, are employed.

Even without pulse-shaping, however, these estimates suggest viable entanglement regimes when $g/A_J$ is larger than about 3. 
In Fig.~\ref{fig:iswap}(b) we plot this ratio using the computed $g$ and $A_J$ from Fig.~\ref{fig:fcigAj} and observe operating bias regions with sufficient ratios of $g / A_J$ within the XRX regime for zero detuning.
The values of the $g / A_J$ ratio in these calculations exceed 10 at certain points, probably limited by the bias resolution, although for such fine features higher-order terms may need to be considered when evaluating noise.
Overall, these calculations suggest it is feasible to use XRX for remote entanglement, though they should not be construed as any kind of upper bound on the achievable spin-photon coupling performance in silicon TQDs. 

Indeed, there are several clear technological avenues for improved performance. 
At present, MW resonators in spin-photon experiments have typical values of $\kappa$ in the few-MHz range down to 1 MHz, corresponding to $Q$s $\approx 1,000-5,000$ \cite{mi_coherent_2018,samkharadze_strong_2018,borjans_resonant_2020,landig_coherent_2018}. 
For $g = 10$~MHz this sets $g / \kappa \approx 10$. Superconducting resonators with $Q > 10^{5}$-$10^{6}$ are now routinely made in other contexts using optimized material processing and packaging \cite{zmuidzinas_superconducting_2012}, and the integration of such techniques with semiconductor qubit technologies promises corresponding improvements in cavity loss, which may allow $g / \kappa$ to approach $10^{3}$. 
On the qubit side, novel techniques or designs that reduce the device $1/f$ noise (\textit{e.g.}, by suppressing the magnitude of charge fluctuations $A_{\mu}$) would directly boost performance by decreasing $A_J$. 
The qubit and cavity design also impact the spin-photon coupling rate $g$; changing the design of the cavity-coupled gate electrode can significantly boost the transverse lever arm beyond that possible via an ordinary plunger gate (as has been assumed in the calculations of Fig. \ref{fig:iswap}). 
For instance, using a higher capacitance split-slit gate design \cite{borjans_split-gate_2020} for cavity coupling would increase the values of $g$ calculated here by about a factor of 2. 
Similarly, cavities using high kinetic inductance superconductors and narrow geometries can boost the impedance into the k$\Omega$ range, increasing the photon voltage \cite{samkharadze_high-kinetic-inductance_2016}; this would raise $g$ by another factor of 4 or more compared to a baseline 50 $\Omega$ resonator, since the coupling scales as $\sqrt{Z_0}$.
These considerations suggest that the $g / A_J$ ratios in Fig. \ref{fig:iswap}(b) are improvable by an order of magnitude or beyond even at present levels of device charge noise, increasing both the operating window in bias space and achievable fidelity. 
As $g$, $A_J$, and $\kappa$ are further optimized, other effects not considered here may also impact the fidelity, such as excitation into higher valley and orbital states during the transition between the TQD qubit and XRX states.

While the calculations here have focused on transverse three-electron dipolar coupling, our methodology naturally extends to considering higher-order couplings, qubits at higher electron number, and longitudinal effects, each of which may be of interest for novel approaches to dot-photon coupling \cite{friesen_decoherence-free_2017,russ_quadrupolar_2018,harvey_coupling_2018} and/or readout \cite{ruskov_quantum-limited_2019}. 
Numerical simulation of more realistic wave functions and device operation may offer new quantitative guidelines or operating insights for these applications as well.

\section{\label{conclude}Conclusion}
Our calculations demonstrate the importance of accurate device modeling for understanding exchange operation, as even qualitative features of simultaneous exchange regimes like RX and ARX turn out to depend strongly on geometry, biasing, and material properties. 
In particular, in silicon TQDs the widest region for strong spin-photon coupling may lie in DQD-like ``XRX'' rather than conventional TQD RX or ARX biasing, as the relatively large effective mass of silicon compared to GaAs requires that electrons be brought closer together for large $J$. 
We estimate that entangling protocol fidelities of 90-99\% are presently within reach for the type of devices we consider and identify several engineering avenues for further improving performance. 

Our results are based on exact calculations of exchange using FCI, which reveal important physical effects not evident in simpler models. 
However, the relatively high computational cost and complexity of these calculations also underscore the importance of developing better approximate models of exchange, including extended Fermi-Hubbard models. 
Such developments will be useful for further exploring device design tradeoffs and studying more sophisticated models of spin-photon coupling in these systems.

\ack
The authors thank C. Anderson, A. A. Kiselev, J. Z. Blumoff, M. Friesen, and C. Tahan for helpful discussions.

\appendix
\section{\label{gcalc}Derivation of Coupling Rate $g$ in QDs}
The dot-photon coupling $g$ is frequently evaluated in circuit QED contexts and is presented in several equivalent forms in the literature, \textit{e.g.}, \cite{childress_mesoscopic_2004,srinivasa_entangling_2016}. For completeness, we rederive it in a form suitable for evaluation with either FH parameters or microscopic wave functions computed from FCI. 

Consider a general Hamiltonian dependent on some time-dependent parameter vector $\vec{V}(t)$, $H(t)=H[\vec{V}(t)]$.  Let $\ket{0}$ and $\ket{1}$ be the two lowest ``instantaneous" eigenstates of this Hamiltonian at time $t$ with ``instantaneous" eigenenergies $hf_1(t)$ and $hf_2(t)$.  In the basis of these time-dependent eigenstates and their frequency difference $f_q(t) = f_1(t) - f_0(t)$, the Hamiltonian is
\begin{equation}
H(t) = hf_{q}(t)\frac{\sigma_z}{2} + \frac{1}{hf_q(t)}\left[
\ket{1}\!\!\bra{1}\frac{dH}{dt}\ket{0}\!\!\bra{0}+ \textrm{h.c.}\right].
\label{diabatic_hamiltonian}
\end{equation}
For dot-photon coupling, the oscillating interaction is generally the sinusoidal voltage of an LC resonator $\mathcal{V}=-iV_0a^\dag\exp(i\omega_rt)+\textrm{h.c.}$, where $a^\dag$ is a microwave photon creation operator.  Here $V_0$ is the root-mean-square (RMS) fluctuation of $\mathcal{V}$; \textit{i.e.}, for any coherent state $\ket{\alpha}$, $V_0= \sqrt{\bra{\alpha}\mathcal{V}^2\ket{\alpha}-\bra{\alpha}\mathcal{V}\ket{\alpha}^2}.$
It is thus interpreted as the RMS voltage of a vacuum fluctuation in the resonator.
By the chain rule, then,
\begin{equation}
\frac{dH}{dt}= 
\frac{\partial H}{\partial V_r} \omega_r V_0 a^\dag\exp(i\omega_rt)+\textrm{h.c.}.
\end{equation}
We thus render Eq.~\ref{diabatic_hamiltonian} as a Jaynes-Cummings Hamiltonian $H=hf_q (\sigma^z/2) + hf_r a^\dag a + hg(a^\dag\sigma^-+a\sigma^+)$ in the rotating frame of the resonator. 
We therefore deduce
\begin{equation}
hg = \frac{f_r}{f_q(t)}\braket{1 |
 H'_r | 0} V_0 .
\end{equation}
When operating close to resonance, $f_r \approx f_q(t)$ and the first term goes to unity, leading to Eq. \ref{eq:hg} in the text.  

The derivation above is independent of geometry; geometric considerations reside in the value of $V_0$.  For a 1-D transmission line resonator, the RMS voltage of mode $k$ is equal to $f_{r,k}\sqrt{{2h Z_0}/{k}}$ \cite{blais_cavity_2004}. The $g$ calculations in the text consider coupling to the fundamental mode $k=1$.

\section{\label{gatenoise}Modeling Gate-Referred Charge Noise}
We consider how the qubit energy splitting $hJ$ depends on charge noise. 
To keep the problem tractable, we focus on fluctuations of the QW chemical potential $\mu_k$ at key positions or ``natural coordinates'' that modulate exchange, such as the potential minima and maxima that set the dot positions and tunnel barrier heights, respectively \cite{reed_reduced_2016}. 
We further assume that these fluctuations, denoted $\delta \mu_k(t)$, are independent and identically distributed (i.i.d.) for each such position $k$, regardless of the device geometry. 
Since we seek to model the noise in terms of gate-referred quantities, we will write
\be
\delta J(t) = \sum_{i,j} \frac{\partial J}{\partial V_i}\frac{\partial V_i}{\partial \mu_j}\delta \mu_j(t) .
\ee
We define the gate lever arm matrix as $\alpha_{km}={\partial \mu_k}/{\partial V_m}$, and switching to vector notation in which $(\nabla_V J)_i = {\partial J}/{\partial V_i}$,
\be
\delta J(t) = (\nabla_V J)\cdot \hat\alpha^{-1}\cdot \delta\vec{\mu}(t).
\ee
The temporal exchange autocorrelation function is then
\be
\label{eq:jautocorr}
\langle \delta J(t) \delta J(t')\rangle  = 
[(\nabla_V J)\cdot (\hat\alpha^\dag\hat\alpha)^{-1} \cdot (\nabla_V J)]
\langle\delta \mu(t)\delta \mu(t')\rangle,
\ee
where we have dropped the subscript on $\delta \mu_k$ and summed the contributions under the assumption of i.i.d. fluctuations in $\mu$.

For TQDs, we are concerned with the five control gates (both P and X gates) and the chemical potentials underneath them. 
(Note that this analysis could be generalized using singular value decomposition to cases where $\alpha$ is not square, \textit{e.g.}, when there are more or fewer gates than relevant coordinates $\mu_k$.)
This implies the noise spectra of the chemical potential and exchange are related by the quantity in brackets in Eq. \ref{eq:jautocorr}. 
This analysis clearly depends on a number of approximations, but has the advantage that the final parameters (the $\alpha$ matrix and the derivatives of $J$) can be obtained from electrostatic modeling or experimental measurements. 
For the calculations presented here we drop the off-diagonal (cross-capacitance) contributions for simplicity, leading to the simpler form of Eq. \ref{eq:AJ}.

\section{\label{multipoles}Exchange Multipoles in XRX}

\begin{figure}
	\includegraphics[width=0.9\textwidth]{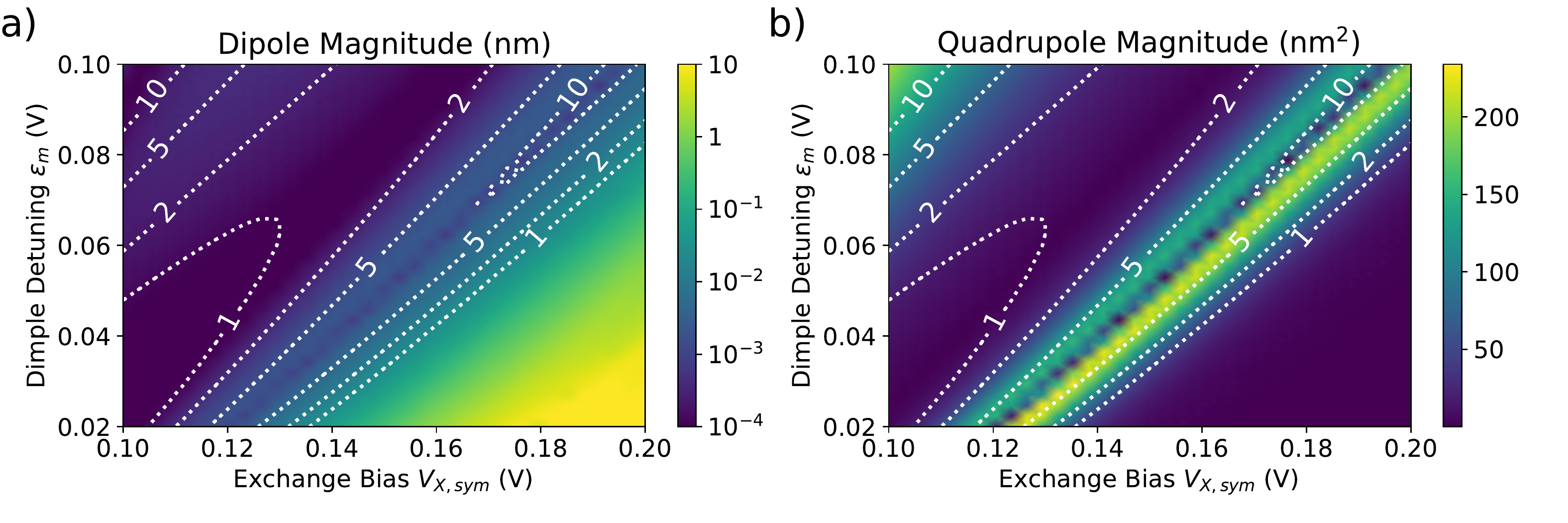}
	\centering
	\caption{Magnitude of (a) dipole ($\sqrt{p_x^2+p_y^2+p_z^2}$) and (b) diagonal quadrupole ($\sqrt{Q_{xx}^2 + Q_{yy}^2 + Q_{zz}^2}$) moments for the FCI charge densities computed in Fig. \ref{fig:fcigAj}. Contours of exchange between 1-10 GHz are overlaid.}
	\label{fig:dipolequad}
\end{figure}

To first order in perturbation theory, the change in energy splitting $hJ$ due to a quasi-static potential fluctuation $\delta V(\vec{r},t)$ is given by
\begin{equation}
h\delta J(t) = \int \delta n_J(\vec{r}) \delta V(\vec{r},t) d\vec{r}
\end{equation}
and is determined by the spatial differential charge density of the qubit eigenstates $\delta n_J(\vec{r}) = n_1(\vec{r}) - n_0(\vec{r})$. 
This suggests we examine the multipole expansion of $\delta n_J(\vec{r})$, which naturally corresponds to the DC charge moment of the qubit. 
The vector and tensor components of the DC dipole and quadrupole moments $p_i$ and $Q_{ij}$, respectively, can be evaluated as
\begin{eqnarray}
p_i &= \int r_i \delta n_J(\vec{r}) d\vec{r} \\
Q_{ij} &= \int \left( 3 r_i r_j - r^{2} \delta_{ij} \right) \delta n_J(\vec{r})d\vec{r}
\end{eqnarray}
where the indices $i,j$ range over the Cartesian coordinates $x,y,z$.

For illustrative purposes in Fig. \ref{fig:dipolequad} we plot the magnitudes of the diagonal components of the qubit dipole and quadrupole for the range of FCI states computed in Fig. \ref{fig:fcigAj}. 
For the traceless quadrupole tensor we show the sum of squares of the diagonal elements $\sqrt{Q_{xx}^2 + Q_{yy}^2 + Q_{zz}^2}$ because our calculations show these terms give the largest contributions to dephasing. 
Intriguingly, we observe that the dipole magnitude is generally very small (of order 0.1 nm and below) at all relevant qubit energies, as expected given the spatial symmetry of zero detuning states. 
We also observe a sharp suppression of the quadrupole moment at the XRX crease corresponding to the narrow ``sweet spot'' observed in the exchange noise calculation of Fig. \ref{fig:fcigAj}. 
It suggests that the sweet spot feature is relatively independent of details of the charge noise model, at least to lowest order in fluctuations.
It is interesting that the three-electron DQD state in XRX suppresses both the DC dipole and quadrupole moments at the sweet spot, in contrast to one-electron DQD charge qubits where the zero detuning sweet spot only cancels out the dipole moment. 
Since the spatial symmetries of the 1e and 3e DQD states are the same (\textit{i.e.}, this is not a TQD ``quadrupole qubit'' \cite{friesen_decoherence-free_2017}), this suggests that the higher order charge moment suppression is due to the screening effect of the additional electrons, which rearrange themselves slightly at the sweet spot to minimize the charge moment. 
However, the narrowness of this feature implies that this cancellation is not robust against small tuning variations, which is related to the observation that RX states are fairly sensitive to higher order noise terms \cite{russ_coupling_2016}.

\section{\label{detuning}Dependence of Simultaneous Exchange on Detuning}
\begin{figure}
	\includegraphics[width=0.9\textwidth]{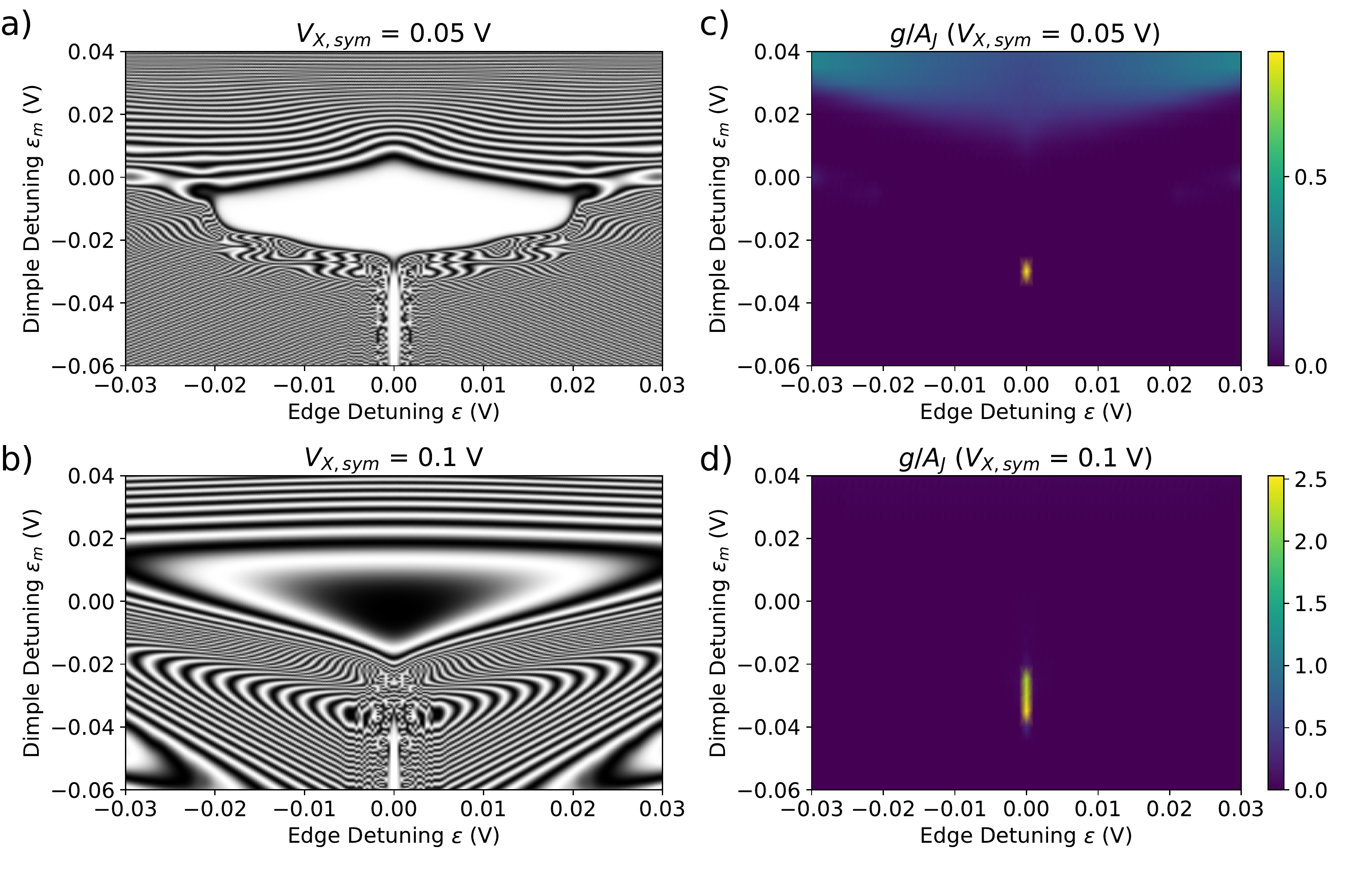}
	\centering
	\caption{FCI calculations of Rabi fringes for (a) $\VXsym = 0.05$ V and (b) $\VXsym = 0.1$ V as a function of dimple and detuning.
	(c)-(d) Corresponding ratios of $g / A_J$ at the same bias conditions.
	}
	\label{fig:detuning}
\end{figure}
In the main text we focus on operation at zero detuning $\epsilon = V_{P1} - V_{P3}$ since charge noise is generally minimized under this condition \cite{russ_coupling_2016,reed_reduced_2016}. 
However, it is useful to observe how exchange varies as a function of $\epsilon$ in FCI calculations. 
In Fig. \ref{fig:detuning}, we show how Rabi fringes and $g / A_J$ evolve with increasing $\VXsym$ as a function of $\epsilon_m$ and $\epsilon$. 
Exchange increases rapidly away from zero $\epsilon$ for both positive and negative dimple, when multiple electrons occupy the same dot. 
As $\VXsym$ increases, we observe a smoothing of exchange fringes at positive dimple where the (111)-(120)-(021) charge boundary would be predicted in FH models. 
These are again due to the smoothness of electrostatic confinement in real devices, which allows electrons to shift position and blurs distinctions between charge states, particularly as X gates are strongly forward biased. 
In FCI calculations this region corresponds to formation of a flat potential underneath X1, P2, and X2, such that detuning mostly translates the electrons laterally underneath the X gates while maintaining roughly the same inter-electron separation, leading to a weak dependence of $J$ on $\epsilon$.
This is another example of the impact of device electrostatics at large $J$.
Fig. \ref{fig:detuning}(c)-(d) also show that the spin-photon coupling ratio $g / A_J$ is maximized near the XRX point at zero detuning, as finite $\epsilon$ introduces asymmetry that increases the qubit dephasing.
We do not see signs of large spin-photon coupling at positive dimple in the bias range shown here, \textit{e.g.}, the (111)-(120)-(021) hybridized state is not apparent. 
Like the negative dimple (111)-(201)-(102) state, this ARX condition is also relatively fragile in bias space for Si/SiGe devices and is only seen under particular bias conditions.

\section{\label{iSWAP}Fidelity of Cavity iSWAP Entanglement}
The cavity iSWAP protocol fidelity can be estimated both analytically and numerically as a function of device parameters. 
We model the time evolution of the coupled-qubit-cavity system using the zero-temperature, rotating-frame master equation
\begin{equation}\eqalign{
	\frac{1}{2\pi}\frac{d\rho}{dt}=&-i\biggl[\sum_j \Delta_j(t)\frac{\sigma^z_j}{2}
	+ g_j(t) (a\sigma_j^++a^\dag \sigma_j^-),\rho\biggr]
	\\
	&-\frac{\kappa}{2}(\{a^\dag a,\rho\}-2a^\dag\rho a)
	-\frac{\gamma}{2}\sum_j  
	(\{\sigma_j^+\sigma_j^-,\rho\}-2\sigma_j^+\rho \sigma_j^-),
}\label{master}\end{equation}
where $\sigma^\pm_j$ are raising and lowering operators and $\sigma^z_j$ Pauli-$Z$ operators for two generic qubits with $j=A,B$, $a$ is the annihilation operator for the cavity, $\gamma$ is a possible qubit relaxation rate in units of Hz, and $\Delta_{A,B}(t)=J_{A,B}(t)-f_r$ is a randomly fluctuating qubit-cavity detuning due to charge noise.   

A detailed simulation would include the finite qubit ramping times; to avoid the complexity of tracking phase-shifts from detuning pulses, we consider ideal pulses which ramp instantaneously from infinite to zero detuning with the resonator, leading to square-pulse modulation of $g_j(t)$.  We integrate from $t=0$ to $t=3/8g$, with $g_A(t)$ set to $g$ for $t<1/8g$ and $0$ for $t>1/8g$.  Likewise, $g_B(t)$ is set to 0 for $t<1/8g$ and to $g$ for $t>1/8g$.  The starting density operator is $\rho(0) = \ket{\psi_0}\!\!\bra{\psi_0}$ with $\ket{\psi_0}=\ket{1}_A\otimes\ket{0}_B\otimes\ket{\textrm{vacuum}}_{\textrm{\tiny cavity}}$.  We have performed simulations comparing 
\begin{itemize}
\item quasistatic noise, where $\gamma=0$ and each $\Delta_j(t)$ is constant but randomly chosen from uncorrelated normal distributions with standard deviation $\sigma$; 
\item $1/f$ noise, in which a time-dependent $\Delta_j(t)$ with $1/f$ power spectral density $S_J(f)=A_J^2/f$ is included in a numeric integrator, sampled using the Voss-McCartney algorithm, again with $\gamma=0$; and 
\item effectively white noise including relaxation, in which $\Delta_j(t)=0$ but $\gamma$ is finite.  
\end{itemize}
The fidelity is then found as $F=\langle\Tr_{\textrm{\tiny cavity}}[\bra{\Psi^-_{AB}}\rho(3/8g)\ket{\Psi^-_{AB}}]\rangle$, where $\ket{\Psi^-_{AB}}$ is the two-qubit antisymmetric Bell state and $\langle\cdot\rangle$ refers to averaging over random instances of $\Delta_j(t)$. 
In this model, the effects of white noise characterized by $\gamma$ and low-frequency detuning fluctuations characterized by $\sigma$ or $A_Z$ are quite different; $F$ is well approximated by $\exp(-\kappa/2 g-\gamma/g)$ in the white noise case and by Eq.~\ref{cqed_analytic} in the quasistatic or $1/f$ noise cases. 
We see negligible impact of the high-frequency components of $1/f$ noise; the results of these simulations appear identical to those using quasistatic noise, and as such are sensitive to the low-frequency cut-off $f_L$ of the $1/f$ noise simulator.
Results of the integration and the analytic approximation are shown in Fig.~\ref{fig:iswap}(a).

In practice for cavity QED, the qubit noise is extracted from a spectroscopic measurement, such as the phase-shift of a microwave pulse reflected from the cavity. 
In the strong coupling regime, the frequency full-width-half-maximum (FWHM) of one of the Rabi-split spectroscopic peaks at zero detuning ($\Delta_j=0$) is $\gamma+\kappa$, resulting from solving Eq.~\ref{master} or standard input-output theory. 
If instead one solves the case of quasistatic noise with $\gamma=0$, the FWHM response is approximately $\sqrt{\kappa^2+4\sigma^2}$.  In this way, we may associate an effective $\gamma_{\textrm{\tiny eff}}$ with $2\sigma$, at least at low $\kappa$, such that
\begin{equation}
\gamma_{\textrm{\tiny eff}}\approx 2\sigma \approx 2 A_J \sqrt{\log(g/f_L)}.
\end{equation}
We note that $\gamma_{\textrm{\tiny eff}}$ should not be interpreted as an effective white noise dephasing rate, to be used for instance in Eq. \ref{master}. Instead, for devices dominated by low-frequency noise, this relation allows approximate conversion from a spectroscopically derived $\gamma_{\textrm{\tiny eff}}$ to the parameter $\sigma$, which is most relevant for a generic fidelity estimate, and in turn to the physical $1/f$ noise amplitude $A_J$.

\section*{References}
\bibliographystyle{unsrt}
\bibliography{RX-spin-photon}
\end{document}